\documentclass{sf2a-conf2011}
\usepackage{graphicx}
\usepackage{hyperref}
\usepackage[]{natbib}  
\usepackage[cyr]{aeguill}
\usepackage{epstopdf}

\def\BibTeX{{\rm B\kern-.05em{\sc i\kern-.025em b}\kern-.08em
    T\kern-.1667em\lower.7ex\hbox{E}\kern-.125emX}}
\bibpunct{(}{)}{;}{a}{}{,}  %%%%%%%%%%%%%  A&A bibliography style
\begin{document}

\TitreGlobal{SF2A 2011}

%%-----------------------------------------------------------------
%%      the top matter
%%

\title{A Bayesian Approach to Gravitational Lens Model Selection}

\runningtitle{Bayesian Lens Model Selection }

\author{I. Balm\`es}\address{Laboratoire Univers et Th\'eories (LUTh), UMR~8102 CNRS, Observatoire de Paris, Universit\'e Paris Diderot, 5 place Jules Janssen, 92190 Meudon, France}

%% IF Author3 has the same affiliation than Author1:
\author{P.-S. Corasaniti$^1$}

%% Keep this line, even if the page will be settled afterwards.
\setcounter{page}{237}

%% To make the final index, repeat the authors here, in the format : Surname, Initial(s) 
\index{Balm\`es, I.}
\index{Corasaniti, P.-S.}

%%-----------------------------------------------------------------

\maketitle

%%-----------------------------------------------------------------
%%        The abstract
%% 
%%  Warning!  within the abstract:
%%  - do not use macros. 
%%  - do not use commands like: \cite, \citet, \citep ... etc.

\begin{abstract}
Over the past decade advancements in the understanding of several
astrophysical phenomena have allowed us to infer a concordance
cosmological model that successfully accounts for most of the
observations of our universe. This has opened up the way to studies
that aim to better determine the constants of the model and confront
its predictions with those of competing scenarios. Here, we use strong
gravitational lenses as cosmological probes. Strong lensing, as
opposed to weak lensing, produces multiple images of a single
source. Extracting cosmologically relevant information requires
accurate modeling of the lens mass distribution, the latter being a
galaxy or a cluster. To this purpose a variety of models are
available, but it is hard to distinguish between them, as the choice
is mostly guided by the quality of the fit to the data without
accounting for the number of additional parameters introduced. 
However, this is a model selection problem rather than one of
parameter fitting that we address in the Bayesian framework. 
Using simple test cases, we show that the assumption of more
complicate lens models may not be justified given the level 
of accuracy of the data.
\end{abstract}

%% Insert the keywords (to appear in the ADS indexing)
%% Keywords must be separated by a comma

%%-----------------------------------------------------------------

\section{Introduction}

Over the past years, our understanding of the universe has greatly
improved. The concordance model explains most of the cosmological
observations. We have now entered a phase where finding new
observational ways of measuring the constants of this model as well as
confronting predictions with those of competing scenarios is crucial
to further advance the research in cosmology. Strongly lensed quasars
constitute such a cosmological probe.

In a strong gravitational lens, each image is the result of a different light-path. As a result, if the source behind the lens has a variable luminosity, as quasars do, it will manifest with a time delay between the two images.

This time delay $\Delta t$ depends on the gravitational potential of the lens, and the underlying cosmological model.

     The time delay between two images A and B is:
         \begin{equation}  
        \Delta t_{A,B} = (1 + z_l) \frac{d_l d_s}{d_{ls}} \left( \frac{1}{2}((\theta_A - \beta)^2 -(\theta_B -\beta)^2) + \psi(\theta_A) - \psi(\theta_B) \right)
        \end{equation}   
        where $\Delta t_{A,B}$, $z_l$, $\theta_A$ and $\theta_B$ are observables, $\beta$, $\psi(\theta_A)$ and $\psi(\theta_B)$ depend on the lens model and $d_l$, $d_s$ and $d_{ls}$ depend on cosmology.
             
         Using the above relation, we can derive constraints on cosmological parameters, provided we assume a lens model. Time delays are particularly sensitive to the value of the Hubble constant $H_0$.

        Unfortunately, a change in the lens model can shift the inferred value of $H_0$ by a factor of two. Hence, the modeling of the lens, as well as a robust discrimination between lens models, is  critical to the study of time delays.
        
        Here, we first discuss the lens models used in our study, then
        describe our methodology based on Bayesian statistical analysis, and finally present our results.

\section{Numerous Lens Models}
      A large number of lens models have been proposed in a vast literature. Given the fact that observables are limited to the position of the images, their time delay and their flux ratio (or magnification), we restrict our analysis to simple examples characterized by a few parameters. In particular we consider two models for lenses with two images, so called ``double'' lenses (for a review on lensing, see~\cite{lens}).

      \begin{enumerate}
        \item{Power-law model:} assume a density profile $\varrho \propto r^{-n}$, with $n$ a free parameter. For $n=2$, it describes an isothermal lens. In order to assess the dependance on the prior parameter interval we assume two different priors: $0 < n < 3$ (large) and $1 < n < 3$ (small).

        \item{Power-law model with external shear:} assume the previous model with the addition of shear accounting for environmental effect on the lens. This adds two parameters: the strength of the shear $\gamma$, and its direction. Expected values for the shear vary up to $\gamma \simeq 0.1$, therefore we assume three different priors on $\gamma$: $\gamma < 0.1$, $< 0.2$ and $< 0.5$ respectively. This allow us to test the shear strength up to nearly unrealistic values.  
      \end{enumerate}

\section{Our method}

To discriminate between different models, we use the Bayes' factor. The reader can find a more complete review on that subject in~\cite{stats}, but it seems useful here to remind a few facts on Bayesian analysis.

Bayesian statistical analysis derives from Bayes' theorem:
\begin{equation}
P(A|B,I)=\frac{P(B|A,I)P(A|I)}{P(B|I)}
\end{equation}

A well known application of this theorem is parameter estimation. If
we take $A=\{\theta\}$ to be a set of parameters in a model, and $B=D$
data resulting from an experience or an observation, then the Bayes'
theorem tells us how our prior knowledge on the parameters $P(\{\theta\}|I)$ is transformed into a new posterior $P(\{\theta\}|D,I)$ by the likelihood $P(D|\{\theta\},I)$. Here $I$ can be written $M_0$, and represents a particular model as well as general background information: before this particular observation, we already had certain expectations about the parameters $A$, coming from our knowledge of the physical world, or from the model we are trying to fit. For example, we might expect a mass to be positive. The resulting equation can be written:
\begin{equation}
\label{paramestim}
P(\{\theta\}|D,M_0)=\frac{P(D|\{\theta\},M_0)P(\{\theta\}|M_0)}{P(D|M_0)}
\end{equation}

But consider another alternative. If we take $A$ to represent a certain model $M_0$, and $B$ to represent our data set, we know have:
\begin{equation}
P(M_0|D,I)=\frac{P(D|M_0,I)P(M_0,I)}{P(D|I)}%=\frac{P(D|\{\theta\},M_0,I)P(\{\theta\}|M_0,I)}{P(\{\theta\}|D,M_0,I)}\frac{P(M_0|I)}{P(D|I)}
\end{equation}
The term $P(D|M_0,I)$ can be calculated from the previous
 equation~\ref{paramestim}: it is the denominator of the
 right-hand-side. $P(M_0|I)$ is our prior belief on the model $M_0$ to
 provide the correct description of the data. $P(D|I)$ could be problematic, but we can get rid of it by considering two models, and studying their relative probability:
\begin{equation}
\frac{P(M_0|D,I)}{P(M_1|D,I)}=\frac{P(D|M_0,I)}{P(D|M_1,I)}\frac{P(M_0|I)}{P(M_1|I)}
\end{equation}

The first term on the right-hand-side of this equation,
$P(D|M_0,I)/P(D|M_1,I)$ is what is called the Bayes' factor
between models $M_0$ and $M_1$. Supposing we have the same prior
belief on the two models, it represents their relative probability,
and quantify the hability of one model to account for the
observations, with respect to the other model.
%    We obtain the likelihood by the usual formula:
 %   \[P(D|\theta,M_i)=\frac{1}{\sqrt{2\pi}\sigma} \exp -\frac{(D_{th}(\theta,M_i)-D)^2}{2\sigma^2} \]

An important advantage of the Bayes' factor over other methods such as
$\chi^2$ or information criterion, is that the Bayes' factor takes
into account a term akin to the Occam's razor idea: models with more
parameters, though they usually fit the data better, should not always
be preferred. In terms of model selection the improvement on the
quality of the fit should be discounted by the increased volume of the
prior model parameter space.

In our particular case, we calculated the Bayes' factor for the lens
models (1) and (2), taking as constraints either the time delay
measurement alone, or the time delay combined with flux ratio measurements.

\section{Our results}

\begin{figure}[t]
%  \sidecaption[t]
  \includegraphics[width=\textwidth]{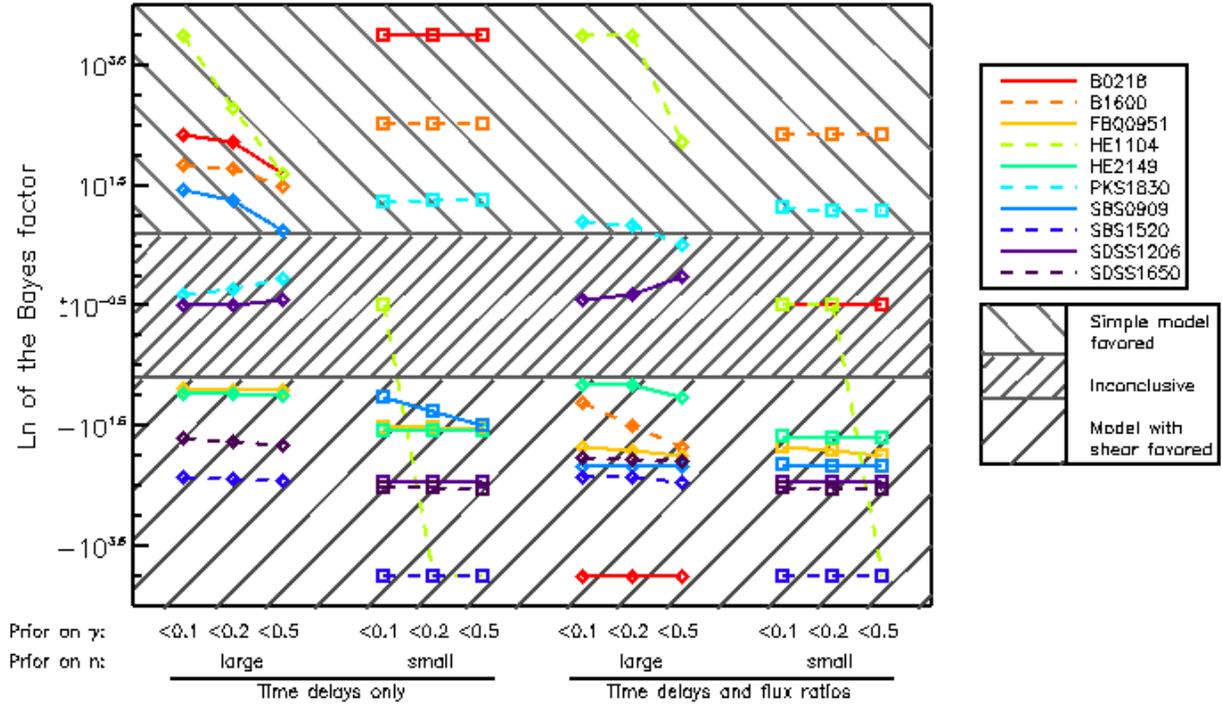}
  \caption{Bayes' factor between model~1 and~2, with different priors. We plot $ln(B_{1,2})$ on a logarithmic scale. Each color represents a different lens}
\label{evidence} 
\end{figure}

Results are summarized in Fig.~\ref{evidence}. Large Bayes' factors favor the simpler model, model~1. Above (below) a certain threshold, the evidence in favor of model~1 (2) is considered strong. In-between, the evidence is inconclusive, and no model can be preferred. In the following, we highlight a few relevant aspects.

    \subsection{Effect of the prior on $n$}
    
    The lens data set is mainly composed of galaxies, which we expect to be nearly isothermal. Nevertheless, our analysis show that few of them are in fact well described by a power-law model with a reduced prior on $n$, while the majority favors the more complex model, which also include the shear: the effects of the environment can not, in general, be taken as negligible. In the case of the larger prior, $0<n<3$, the number of lenses well described by the power-law increases.
    
    \subsection{Effect of the prior on $\gamma$}
    
    In more than half of the cases, allowing higher (unrealistic) shear strength does not change the Bayes' factor. This is a consequence of Occam's razor: as the parameter space grows, the fit gets better and better, but this effect is compensated for by the Bayes' factor.    
    
    In about a quarter of the cases, widening the prior on $\gamma$ favors the more complex model, as the fit gets sufficiently better to over-compensate for the Occam's razor term.
    
    \subsection{Effect of the flux ratios}
    
    Time delays depend on the gravitational potential of the lens, whereas flux ratios depend on its second derivative. Furthermore, they are subject to a number of local phenomena (microlensing, absorption...) that do not affect time delays. Therefore, flux ratios can be hardly described with a smooth model, eventually requiring a more complex modeling than needed by time delays. This is consistent with our findings: in fact, adding flux ratios as a constraint leads to having less lenses accurately described by model 1, since $\ln B_{1,2}$ decreases.

\section{Conclusion}
  
  Bayesian techniques are a good way to discriminate between lens models and allow us to decide which double lenses can be accurately modeled by a simple power-law model. With the result from this analysis, we now have a good sample of double lenses, together with lens models, to determine cosmological parameters more accurately. %Balmes & Corasaniti, in prep ?
    
% Optional acknowledgements
% -------------------------
\begin{acknowledgements}
I.~Balm\`es is supported by a scolarship of the "Minist\'ere de l'\'Education Nationale, de la Recherche et de la Technologie" (MENRT).
\end{acknowledgements}

\bibliographystyle{aa}  % A&A bibliography style file (aa.bst)
\bibliography{balmes} % your references in file: Yourfile.bib

\begin{thebibliography}{2}
\expandafter\ifx\csname natexlab\endcsname\relax\def\natexlab#1{#1}\fi

\bibitem[{{Kochanek}(2006)}]{lens}
{Kochanek}, C.~S. 2006, in Saas-Fee Advanced Course 33: Gravitational Lensing:
  Strong, Weak and Micro, ed. {G.~Meylan, P.~Jetzer, P.~North, P.~Schneider,
  C.~S.~Kochanek, \& J.~Wambsganss}, 91--268

\bibitem[{{Trotta}(2008)}]{stats}
{Trotta}, R. 2008, Contemporary Physics, 49, 71

\end{thebibliography}

\end{document}